\documentclass[apj]{emulateapj}
\pdfoutput=1
\usepackage{hyperref}
\usepackage{amsmath,amstext}
\usepackage[T1]{fontenc}
\usepackage[figure,figure*]{hypcap}

\shorttitle{J0804+2239: A DZ star with CIA}
\shortauthors{Blouin et al.}

\begin{document}

\submitted{Accepted for publication in The Astrophysical Journal}

\title{A New Generation of Cool White Dwarf Atmosphere Models. II. A DZ Star with Collision-Induced Absorption}

\author{S. Blouin\altaffilmark{1}}
\author{P. Dufour\altaffilmark{1}}
\author{N.F. Allard\altaffilmark{2,3}}
\author{M. Kilic\altaffilmark{4}}

\altaffiltext{1}{D\'epartement de Physique, Universit\'e de Montr\'eal, Montr\'ea\
l,
  QC H3C 3J7, Canada; sblouin@astro.umontreal.ca, dufourpa@astro.umontreal.ca.}
\altaffiltext{2}{GEPI, Observatoire de Paris, Universit\'e PSL, CNRS, UMR 8111,
  61 avenue de l'Observatoire, 75014 Paris, France.}
\altaffiltext{3}{Sorbonne Universit\'e, CNRS, UMR 7095,
  Institut d'Astrophysique de Paris, 98bis boulevard Arago, 75014 Paris, France.}
\altaffiltext{4}{Department of Physics and Astronomy, University of Oklahoma,
  440 W. Brooks St., Norman, OK 73019, USA}

\begin{abstract}
  In the first paper of this series \citep{blouin2018model}, we presented our
  upgraded cool white dwarf atmosphere code. In this second paper, we use our new
  models to analyze SDSS J080440.63+223948.6 (J0804+2239), the first DZ star to
  show collision-induced absorption (CIA). 
  This object provides a crucial test for our models, since previous
  versions of our code were unable to simultaneously fit the metal absorption lines
  and the CIA.
  We find an excellent fit to both the spectroscopic and photometric data, which
  further validates the improved constitutive physics of our models.
  We also show that the presence of metal lines allows to lift the degeneracy
  between high and low hydrogen abundances that usually affects the fits of white
  dwarfs with CIA. Finally, we investigate the potential impact of spectroscopically
  undetected metals on the photometric solutions of DC stars.
  \end{abstract}
\keywords{stars: atmospheres --- stars: individual (SDSS J080440.63+223948.6) --- white dwarfs}

\section{Introduction}

Most cool white dwarfs ($T_{\rm eff} \lesssim 5000\,{\rm K}$) are classified as DC stars, meaning that
they have a featureless spectrum. At such temperatures, there is too little thermal energy
to excite the transitions that are required to produce hydrogen or helium spectral lines
in the optical and infrared. Therefore, 
all the information we can get from these objects is limited to photometric observations
and parallaxes.
This is sufficient to obtain the effective temperature
$T_{\rm eff}$ and the surface gravity $\log g$ by fitting the spectral energy distribution
(SED) with atmosphere models
\cite[e.g.,][]{bergeron1997chemical,bergeron2001photometric}. It is even 
possible to deduce the atmospheric composition of these objects through a detailed analysis
of their SED. However, different atmosphere model codes often yield different
results. For instance, \cite{kowalski2006found}, \cite{kilic2009spitzer} and \cite{kilic2009near} 
conclude that there are virtually no helium-rich DC stars below $T_{\rm eff}=5000\,{\rm K}$, while the findings of
\cite{bergeron1997chemical}, \cite{bergeron2001photometric}, \cite{kilic2006cool} and \cite{kilic2010detailed}
suggest that hydrogen and helium-rich stars are roughly equally abundant in this temperature range.
These differences arise mainly because the pure helium atmosphere models of \cite{kowalski2006found} 
are similar to blackbodies, while those of the Montreal group 
\citep{bergeron2001photometric,bergeron1995new} are not.
Since cool white dwarfs do not have blackbody SEDs, using the models of \cite{kowalski2006found}
leads to all cool DC stars being assigned a H-rich composition.

Since the featureless spectra of cool DC white dwarfs provide very few opportunities to
test and compare different atmosphere models, problems with these models have not yet been identified.
In the absence of any observational test, it is hard to say which atmosphere code should be trusted.
Fortunately, cool metal-polluted white dwarfs (DZ stars) show atomic absorption lines that can be used
to diagnose the accuracy of different atmosphere models. Properly fitting the spectral lines of
DZ stars requires models that precisely map the temperature and density conditions of the
line-forming regions of the atmosphere. This is particularly true for cool DZ stars, where spectral
lines can significantly differ from conventional Lorentzian profiles
\citep{allard2018mgi,allard2016mgii,allard2016asymmetry,allard2014caii,allard2014nai}.
Good examples of this deviation
include the \ion{Mg}{2} 2795/2802{\,\AA} lines of van Maanen 2 \citep{wolff2002element} 
and Ross 640 \citep{blouin2018model,koester2000element}, as well as the \ion{Na}{1} D doublet of WD 2356-209 
\citep{bergeron2005interpretation,homeier2007,homeier2005}.

Another way of diagnosing atmosphere models is to examine cool white dwarfs that show collision-induced
absorption (CIA) from molecular hydrogen in the infrared (IR). CIA arises due to collisions between H$_2$
and other particles (H, H$_2$ or He), which lead to the induction of an electric dipole that enables
IR absorption \citep{lenzuni1991rosseland,frommhold1993collision}. The intensity of the IR flux depletion
resulting from CIA depends mainly on the hydrogen abundance and the photospheric density, thus allowing
to constrain the physical conditions in these white dwarfs. Recent theoretical calculations \citep{blouin2017cia}
also show that for white dwarfs with a mixed H/He atmosphere and an effective temperature below $4000\,{\rm K}$, 
the H$_2$-He CIA profiles undergo a density-driven distortion, which could provide another way of assessing
the physical conditions in the atmopsheres of these objects.

In the first paper of this series \citep{blouin2018model}, we presented an improved atmosphere
model code that relies on ab initio calculations to properly model cool DZ stars. We showed how this
upgraded code allows better spectroscopic fits for two cool DZ stars (LP 658-2 and Ross 640) that presented
a challenge to previous atmosphere models. These improved fits suggest that our models properly capture
the physics and chemistry of cool white dwarf atmospheres. To extend the validation of our code, we turn to
SDSS J080440.63+223948.6 (J0804+2239), the first DZ star to show CIA. Originally identified by \cite{kilic2010detailed},
this star has not yet been the object of any published spectroscopic or photometric fit. Since it presents both
metal absorption lines and CIA, this star provides a crucial test for our models. 
Our detailed analysis of J0804+2239 is presented in Section \ref{sec:analysis}, we discuss of the possibility of
parameter degeneracies in Section \ref{sec:degeneracies} and our conclusions are given in
Section \ref{sec:conclusion}.

\section{Analysis of J0804+2239}
\label{sec:analysis}

\subsection{Observations}
\cite{kilic2010detailed} identified J0804+2239 as a DZ white dwarf based on a low-resolution
spectrum from the Hobby-Eberly Telescope. These spectra were obtained with the
Marcario Low Resolution Spectrograph \citep[LRS;][]{hill1998hobby} on UT 2005 October 30
using the G1 grism with a $2.0\arcsec$ slit and the GG385 blocking filter.
This setup provided an $R=300$ optical spectrum redward of 4100{\,\AA}, which revealed
significant absorption features from \ion{Ca}{1} and the Na D doublet. The discovery of both
metal lines in its optical spectrum and a significant flux deficit in the infrared
prompted us to obtain a higher resolution and better quality spectrum.

We used the 6.5-m MMT with the Blue Channel Spectrograph to obtain additional spectroscopy
of J0804+2239 on UT 2009 November 20. We operated the spectrograph with the 500 line mm$^{-1}$
grating in first order, providing wavelength coverage from 3660 to 6800{\,\AA}
and a resolving power of $R=1200$ with the $1.25\arcsec$ slit. We obtained four 3-min exposures and one 2-min
exposure of the target. We obtained all spectra at the parallactic angle and acquired a He--Ar--Ne
comparison lamp exposure for wavelength calibration. We used the observations of the
spectrophotometric standard star G24-9, which is also a cool white dwarf, for flux calibration. 
Figure \ref{fig:bestfit_spectro} shows the MMT spectrum of J0804+2239, 
which reveals additional lines from \ion{Ca}{2} and Fe in the blue.

Our photometric analysis of J0804+2239 is based on the $ugriz$ and $JHK$ photometry reported in 
\cite{kilic2010detailed} and the \textit{Gaia} DR2 parallax \citep{prusti2016gaia,brown2018gaia},
which are given in Table \ref{tab:observations}. Note that this object was also observed by the 
UKIDSS Large Area Survey \citep{lawrence2007ukidss} with photometric measurements that are
consistent with those of \cite{kilic2010detailed} within the errors.

\begin{deluxetable}{cc}
  \tablecaption{Observational data. \label{tab:observations}}
  \tablehead{\colhead{Observation} & \colhead{Value}}
  \startdata
  Parallax (mas)  &  $25.28 \pm 0.14$  \\
  $u$           &  $19.73 \pm 0.03$  \\
  $g$             &  $18.30 \pm 0.03$  \\
  $r$             &  $17.59 \pm 0.03$  \\
  $i$             &  $17.39 \pm 0.03$  \\
  $z$             &  $17.33 \pm 0.03$  \\
  $J$             &  $16.71 \pm 0.04$  \\
  $H$             &  $16.92 \pm 0.04$  \\
  $K$             &  $17.29 \pm 0.06$  
  \enddata
\end{deluxetable}

\subsection{Best fit}
\label{sec:bestfit}

To analyze J0804+2239, we
use a new grid of atmosphere models computed using the code described in the first paper of this series
\citep{blouin2018model}.
This code takes into account numerous high-pressure effects relevant for the modeling of cool DZ stars.
In particular, ab initio equations of state for helium and hydrogen are used \citep{becker2014ab},
the most important spectral lines
(\ion{Ca}{1} 4226\,{\rm \AA}, \ion{Ca}{2} H \& K, \ion{Mg}{1} 2852\,\AA, \ion{Mg}{2} 2795/2802\,\AA,
the Mgb triplet and the \ion{Na}{1} D doublet)
are computed using the unified line shape theory
of \cite{allard1999effect}, the nonideal ionization of helium and metals is taken into account using ab initio
results \citep{kowalski2007equation,blouin2018model}, the high-density distortion of H$_2$-He CIA profiles is 
included \citep{blouin2017cia}, continuum opacities are corrected for collective interactions between
atoms \citep{rohrmann2018rayleigh,iglesias2002density} and the opacity of the far red wing of the Ly$\alpha$ line
is included \citep{kowalski2006found}. 
The model grid we use to fit J0804+2239 extends in four distinct dimensions:
hydrogen abundance (helium only, hydrogen only and $\log\,{\rm H/He}$ from $-5$ to $0$ in steps of 0.5 dex), 
metal abundance ($\log\,{\rm Ca/He}$ from $-12$ to $-8$ in steps of 0.5 dex), effective temperature (from 3500
to 7000\,K in steps of 250\,K) and surface gravity (from 7.0 to 9.0 in steps of 0.5 dex).

We use the photometric technique \citep{bergeron2001photometric} to find $T_{\rm eff}$, $\log g$ and the 
hydrogen abundance of J0804+2239. In particular, the solid angle $\pi (R/D)^2$, $T_{\rm eff}$ and the abundance
ratio H/He are found using a $\chi^2$ minimization technique based on the Levenberg-Marquardt algorithm.
The photometric observations are converted into fluxes using the constants reported in \cite{holberg2006calibration}
and the evolutionary models of \cite{fontaine2001potential} are used to find the mass of the star from its
solid angle and parallax. Once a consistent photometric solution is found, we use the spectroscopic
observations to fit the abundance of Ca, Fe and Na
\footnote{All other heavy elements (from C to Cu) are included in the models, but since we could not
  use any spectral line to fit their abundance, we simply assumed the same abundance ratio with respect to Ca
  as in chondrites \citep{lodders2003solar}.}.
More specifically, we use the \ion{Ca}{2} H \& K and \ion{Ca}{1} 4226{\,\AA} lines to constrain the Ca/He 
ratio, the small spectroscopic features near $3700-3800$\,{\AA} for Fe/He and the Na D doublet for Na/He.
The metal abundance ratios thus found being different from our initial guesses, the photometric loop is 
repeated until internal consistency is reached.

\begin{figure}
  \includegraphics[width=\columnwidth]{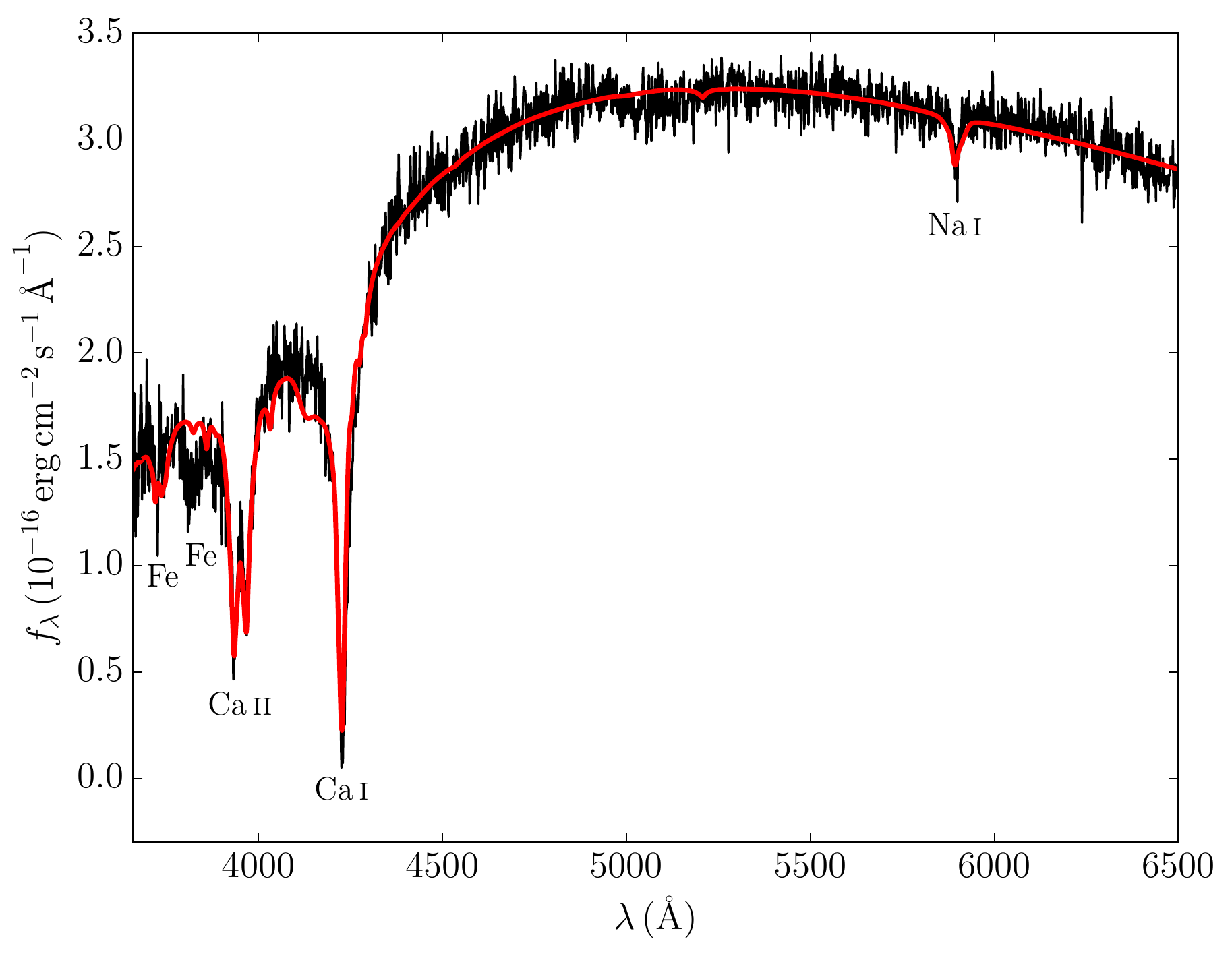}
  \caption{Best spectroscopic solution. The corresponding fitting parameters are given in Table \ref{tab:params}.}
  \label{fig:bestfit_spectro}
\end{figure}

\begin{figure}
  \includegraphics[width=\columnwidth]{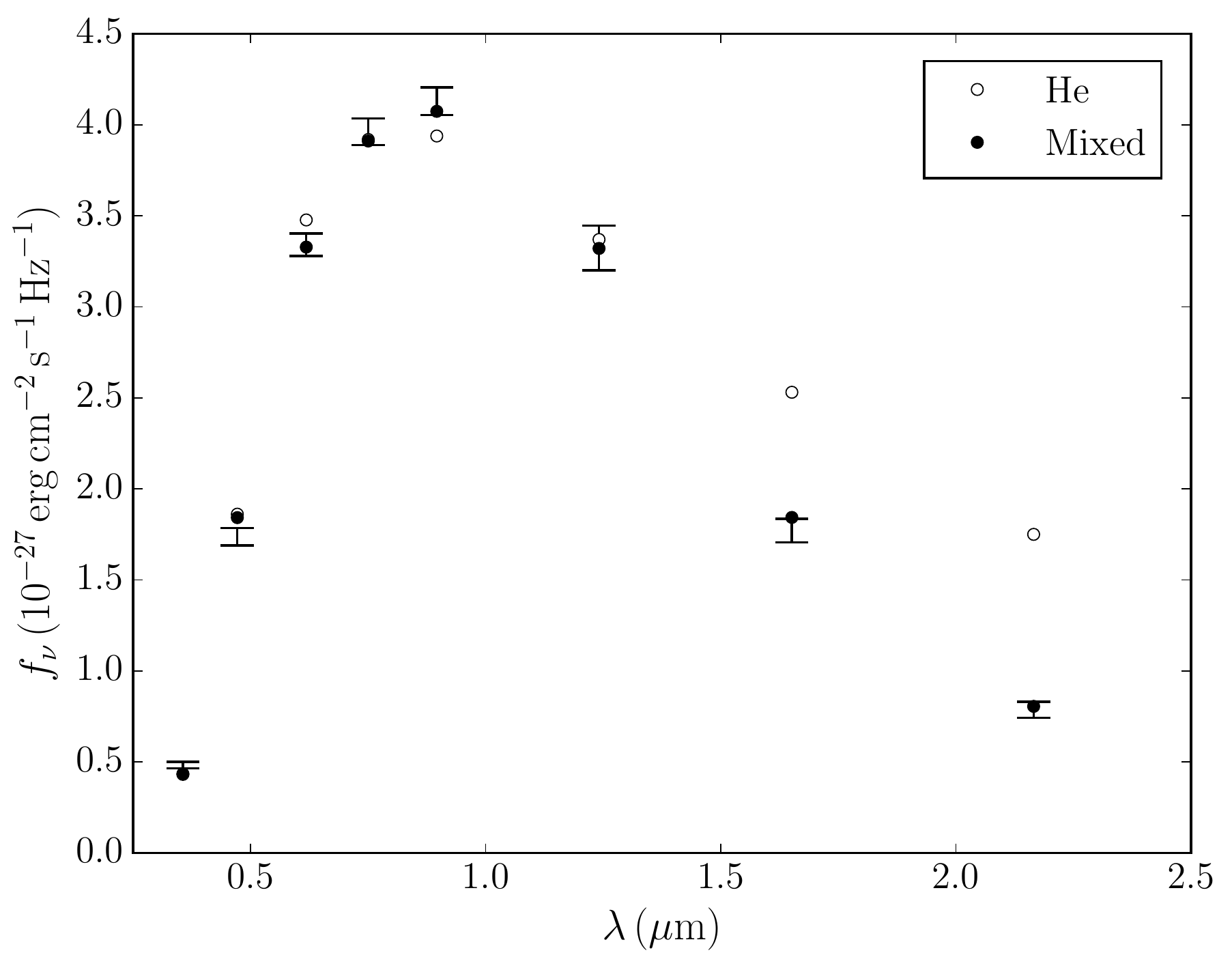}
  \caption{Comparison between the best photometric solution found when assuming a hydrogen-free atmosphere (open circles)
  and the best solution found when the H/He ratio is fitted to the photometric observations (filled circles). The atmospheric
  parameters of the mixed model are given in Table \ref{tab:params} and the hydrogen-free model has 
  $T_{\rm eff}=4620\,{\rm K}$, $\log g =7.93$ and $\log {\rm Ca/He}=-10.5$.}
  \label{fig:ir_deficit}
\end{figure}

The parameters of our best-fit solution are given in Table \ref{tab:params}. The uncertainties were
estimated by manually altering the solution parameters and visually inspecting the resulting fits. 
The corresponding spectroscopic and photometric solutions
are shown in Figures \ref{fig:bestfit_spectro} and \ref{fig:ir_deficit}. The spectroscopic fit is excellent: 
in particular, the \ion{Ca}{2} H \& K lines, the \ion{Ca}{1} 4226\,{\AA} line and
the Na D doublet are all well reproduced by our model.
However, we note that our fit to the weak Fe lines is not so good. This is not surprising insofar as these
lines are still computed assuming Lorentzian profiles, which are known to poorly reproduce the spectral features
of cool DZ stars \citep{allard2018mgi}. Nevertheless, since these Fe lines are weak, they have a very
limited impact on the model atmosphere structure and they do not affect our solution.
As shown in Figure \ref{fig:ir_deficit}, the photometric fit is satisfactory:
all model fluxes are within (or very close to) the observational uncertainties
and the reduced $\chi^2$ is 2.0. 

\begin{deluxetable}{rl}
  \tablecaption{Fitting parameters. \label{tab:params}}
  \tablehead{\colhead{Parameter} & \colhead{Value}}
  \startdata
  $T_{\rm eff}$      & $\phm{-} \;\, \phd 4970 \pm 100\,{\rm K}$ \\
  $\log g$           & $\phm{-} \;\, \phn 7.98 \pm 0.05$ \\
  $\log {\rm H/He}$  & $\phn \phn -1.6 \pm 0.2$ \\
  $\log {\rm Ca/He}$ & $\phn -10.0 \pm 0.1$\\
  $\log {\rm Fe/He}$ & $\phn \phn -9.8 \pm 0.2$\\
  $\log {\rm Na/He}$ & $\phn -11.0 \pm 0.2$\\
  $\tau_{\rm cool}$  & $\phn\phn\phn\;\;\, 5.9 \pm 0.6\,{\rm Gyr}$
  \enddata
\end{deluxetable}

Also shown in Figure \ref{fig:ir_deficit} is the best solution found when assuming a hydrogen-free atmosphere.
To obtain this solution, we used the whole fitting procedure described above to find the best $T_{\rm eff}$, $\log g$ and 
metal abundances if a hydrogen-free atmosphere is assumed. This exercise clearly shows that there must be hydrogen
in the atmosphere of this star, since there is no other way to generate the IR flux depletion observed in the photometric
measurements.

Note that with the previous version of our atmosphere code \citep{dufour2007spectral}
we were unable to find a solution that simultaneously fitted both the photometry and the spectroscopy.
Moreover, the atmospheric parameters of the best solution found with those models were quite
different ($T_{\rm eff}=4780\,{\rm K}$, $\log g = 7.80$, $\log {\rm H/He}=-3$ and $\log {\rm Ca/He}=-10.9$), resulting
in a cooling time of 4.2 Gyr, which is 1.7 Gyr shorter than the cooling time found with our improved models.
The next section describes the improvements made to our model atmosphere code that explain the
difference between both solutions.

\subsection{On the importance of the improved constitutive physics}
\begin{figure}
  \includegraphics[width=\columnwidth]{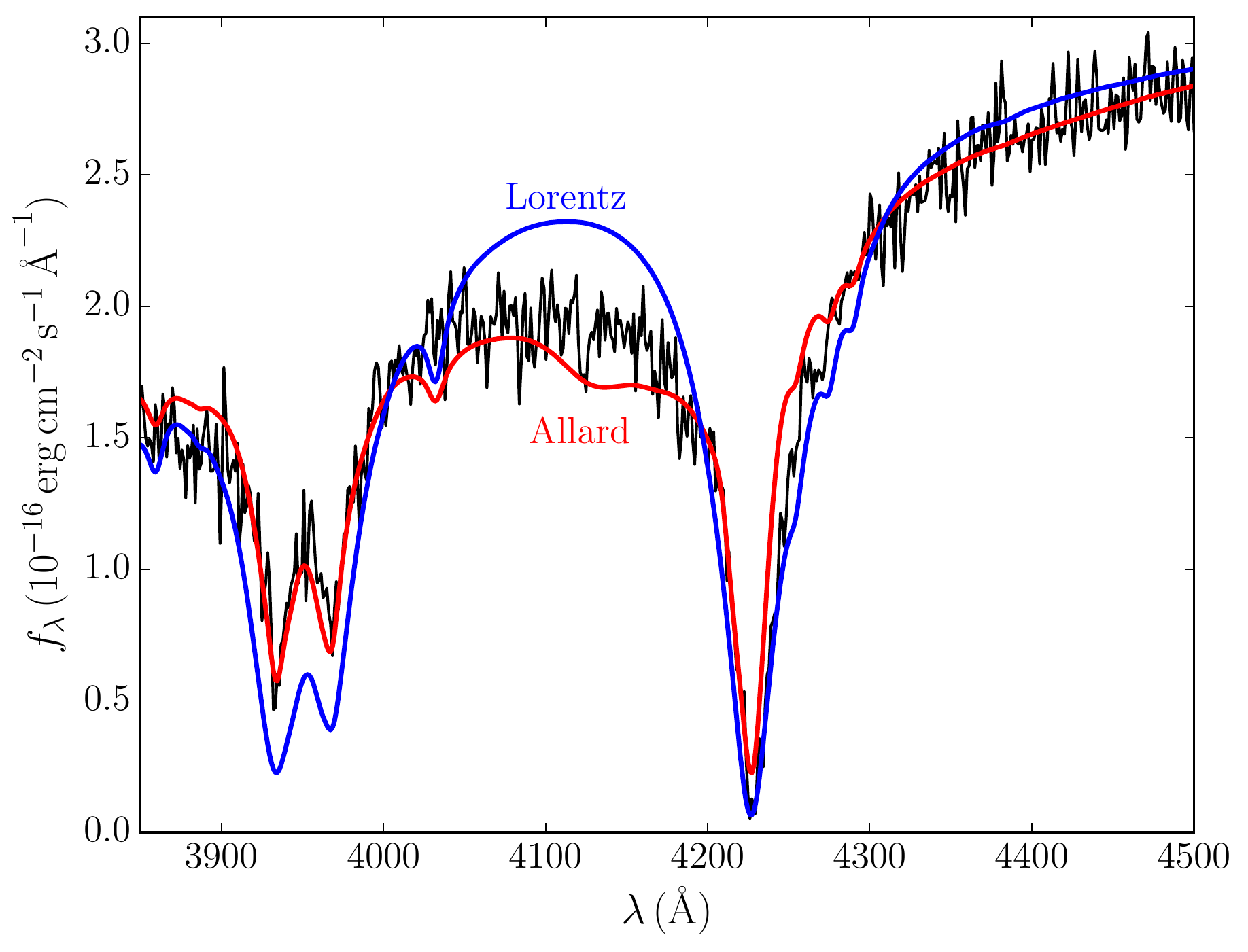}
  \caption{Comparison between the best solutions found when line profiles are computed using 
    the unified line shape theory of \cite{allard1999effect} (in red) and when conventional 
    Lorentzian profiles are assumed (in blue).}
  \label{fig:allardCa}
\end{figure}

The density at the photosphere of our best-fitting model reaches $0.07\,{\rm g\,cm}^{-3}$ 
($n_{\rm He} = 1.05 \times 10^{22}\,{\rm cm}^{-3}$).
This density is low enough that many nonideal high-density effects (e.g., nonideal ionization equilibrium,
\citealt{kowalski2007equation,blouin2018model} and nonideal dissociation equilibrium, \citealt{kowalski2006dissociation})
are negligible. Nevertheless, the photospheric density is high enough to significantly affect spectral line
profiles. Figure \ref{fig:allardCa} compares the best spectroscopic solution found in the previous section to the
best solution found from a model grid computed assuming Lorentzian profiles for all metal lines. Clearly, the improved
profiles for \ion{Ca}{2} H \& K \citep{allard2014caii} and \ion{Ca}{1} 4226\,{\AA} lead to a much better spectral
fit.
Still, our \ion{Ca}{1} 4226\,{\AA} profile is not perfect, since it predicts an unobserved small opacity bump around 4140\,{\AA}.
This could be due to the fact that our \ion{Ca}{1} 4226\,{\AA} profile was derived
from interaction potentials that are not as accurate as the ones used for the rest of our improved line profiles (the Ca-He potentials
were found through open-shell configuration-interaction singles calculations, using
the ROCIS module of the ORCA quantum chemistry package,
\citealt{neese2012orca}). We expect that more accurate potentials will resolve this issue.

\begin{figure}
  \includegraphics[width=\columnwidth]{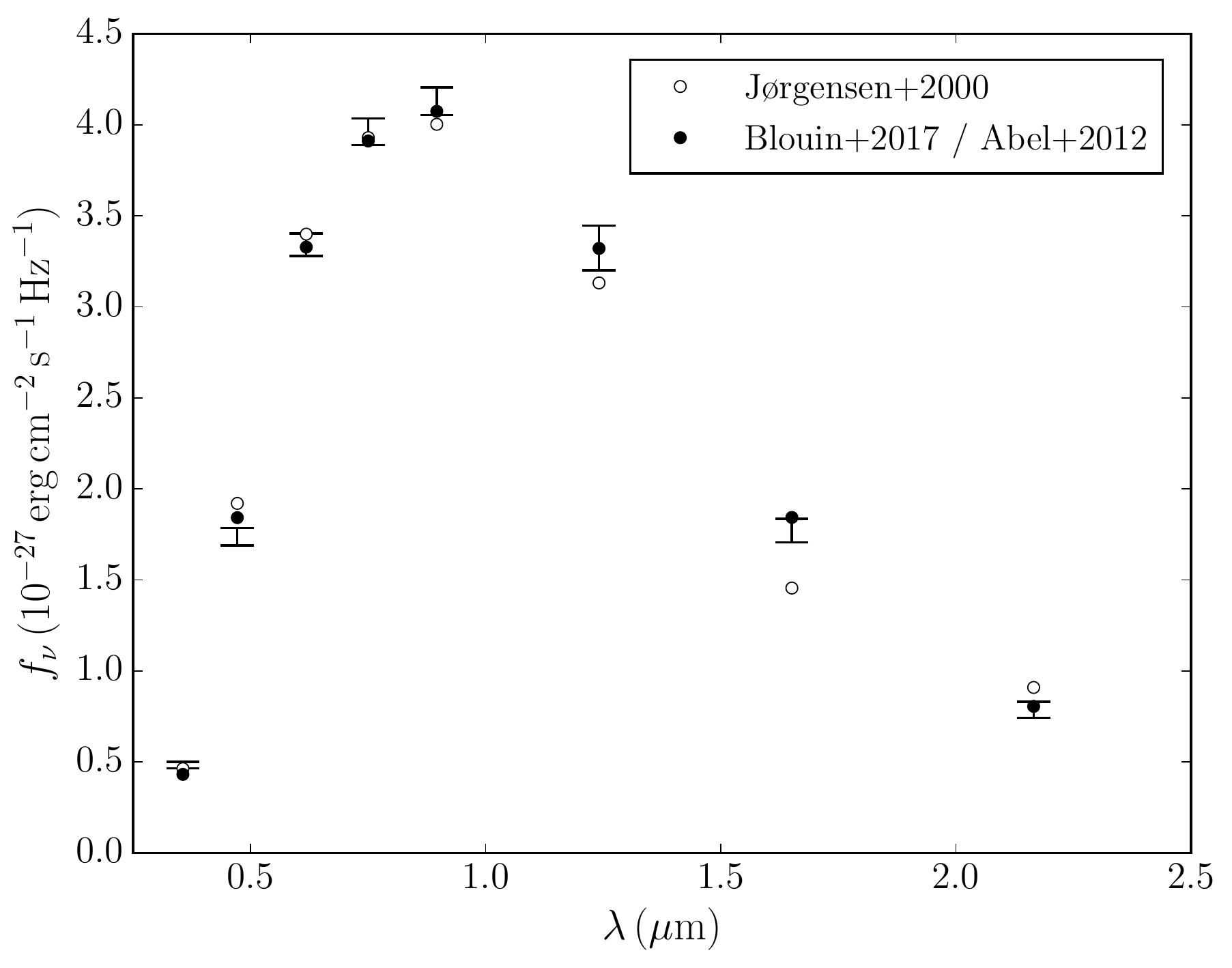}
  \caption{Best photometric solutions for model grids with $\log {\rm H/He}=-1.6$ and different CIA implementations.
    The filled circles show the solution obtained using the CIA profiles of \cite{abel2012infrared} or \cite{blouin2017cia} 
    (which are virtually identical for the physical conditions encountered at the photosphere of J0804+2239) and the open
    circles illustrate the best fit found when using the CIA profiles of \cite{jorgensen2000atmospheres}. $T_{\rm eff}$ 
    and $\log g$ are adjusted to fit the photometry, while spectroscopy is used to fit the metal abundance.}
  \label{fig:cia_model}
\end{figure}

\begin{figure}
  \includegraphics[width=\columnwidth]{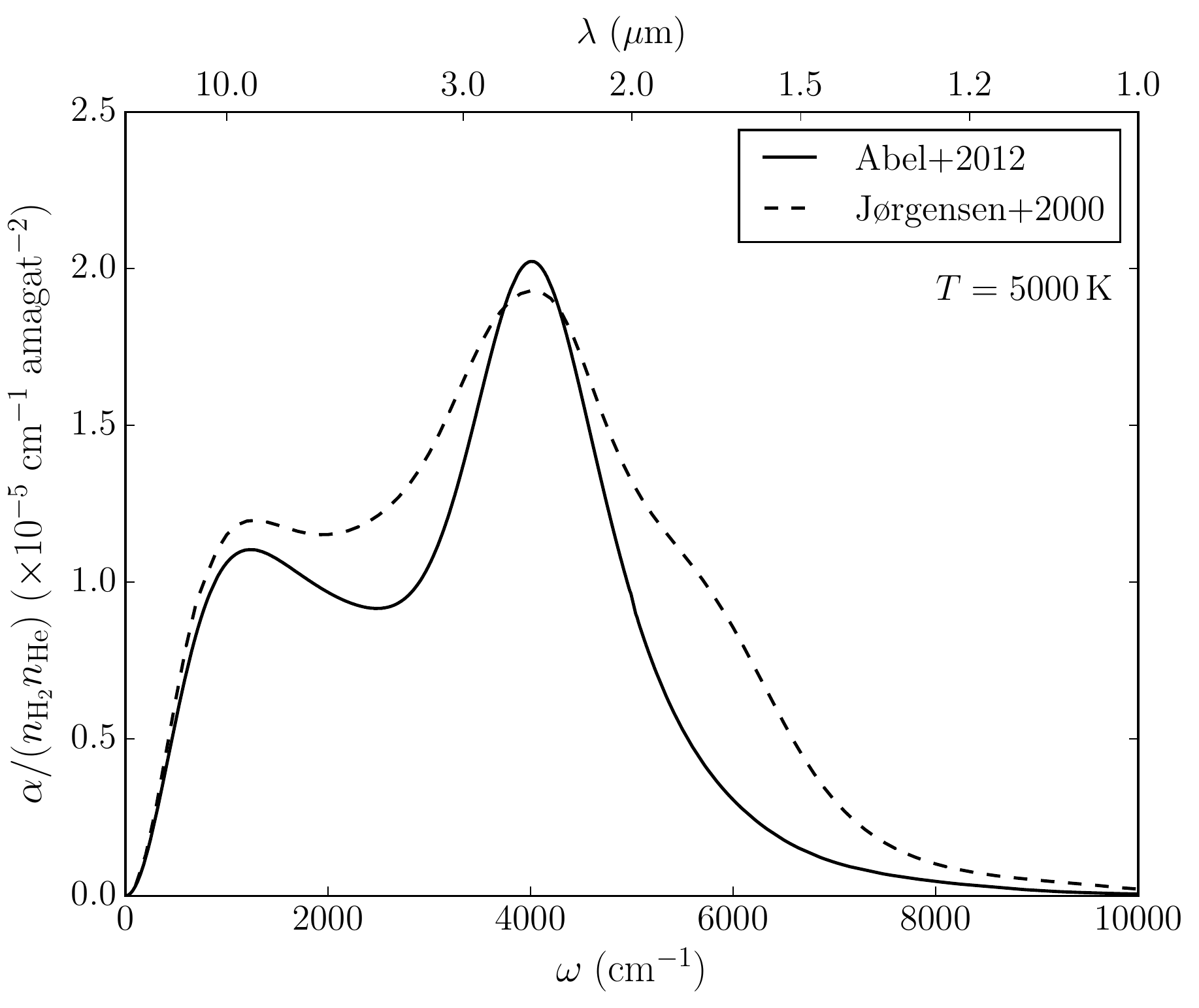}
  \caption{H$_2$-He CIA profiles at 5000\,K, as computed by \cite{abel2012infrared} and \cite{jorgensen2000atmospheres}. 
    The spectra are divided by the number density
  of H$_2$ and He.}
  \label{fig:cia_prof}
\end{figure}

Another improvement to the atmosphere code that proved to be crucial to obtain a good fit to J0804+2239 is the use
of improved CIA profiles. As mentioned above, we use the H$_2$-He CIA profiles given in \cite{blouin2017cia}. They
consist of the profiles computed by \cite{abel2012infrared}, but corrected for various high-density nonideal effects.
Below a photospheric density of about $0.1\,{\rm g\,cm}^{-3}$ 
($n_{\rm He} = 1.5 \times 10^{22}\,{\rm cm}^{-3}$)
, these effects are almost nonexistent and the CIA
profiles used to compute our best-fitting model are therefore virtually identical to those of \cite{abel2012infrared}.
Before the implementation of these profiles, our atmosphere models were relying on the CIA profiles of 
\cite{jorgensen2000atmospheres}. Using these profiles, we found that it is impossible to obtain a photometric 
solution that is as good as the one found using the profiles of \cite{abel2012infrared}. In fact, as shown in
Figure \ref{fig:cia_model}, the profiles of \cite{jorgensen2000atmospheres} lead to an SED that has not enough
flux in the $J$ and $H$ bands to fit the photometric observations. This is a direct consequence of the strong disagreement
between the CIA profiles of \cite{jorgensen2000atmospheres} and \cite{abel2012infrared} in the $\approx 1.2-2\,\mu{\rm m}$
region (see Figure \ref{fig:cia_prof}). We are confident that the profiles of \cite{abel2012infrared} are more accurate,
since \cite{jorgensen2000atmospheres} computed their CIA profiles using potential energy and
induced dipole surfaces that were obtained with a smaller basis set. Moreover, using an entirely independent method
(i.e., molecular dynamics with density functional theory), \cite{blouin2017cia} found absorption spectra that were
in better agreement with \cite{abel2012infrared} than with \cite{jorgensen2000atmospheres} in the $\approx 1.2-2\,\mu{\rm m}$
interval.

\section{Possible degeneracies}
\label{sec:degeneracies}

\subsection{Hydrogen abundance in J0804+2239}
\label{sec:cia_degen}
In our atmosphere models, the intensity of the H$_2$-He CIA reaches a maximum when $\log {\rm H/He} \approx -2.5$. 
This maximum is the result of two competing effects. On one hand, if the H/He ratio is high, H$_2$ is more abundant and 
the H$_2$-He CIA intensity increases. On the other hand, if there is little hydrogen in the atmosphere, the density of 
the atmosphere is higher,
which also leads to a stronger CIA. Because of this maximum, $\chi^2$ minimization algorithms often find two solutions when
fitting the photometric observations of white dwarfs with CIA in their spectrum: one above the $\log {\rm H/He} \approx -2.5$
maximum and one below. In the case of DC stars, the spectroscopic data is of no help to decide between the two solutions, so 
the usual approach is to simply keep the solution that has the smallest $\chi^2$. However, the difference between the two 
solutions can be quite small and minute changes in the models or in the observations can make the difference between choosing
a solution instead of the other. As an example, \cite{kilic2010detailed} concluded that SDSS J030924.87+002525.3,
SDSS J143718.15+415151.5 and SDSS J172257.78+575250.7 have, respectively, $\log\,{\rm H/He}$ abundance ratios of $-4.43$, 
$-4.26$ and $-4.21$, while \cite{gianninas2015ultracool} found that the same three stars have abundance ratios of 
$-1.30$, $-1.97$ and $-1.49$.

\begin{figure}
  \includegraphics[width=\columnwidth]{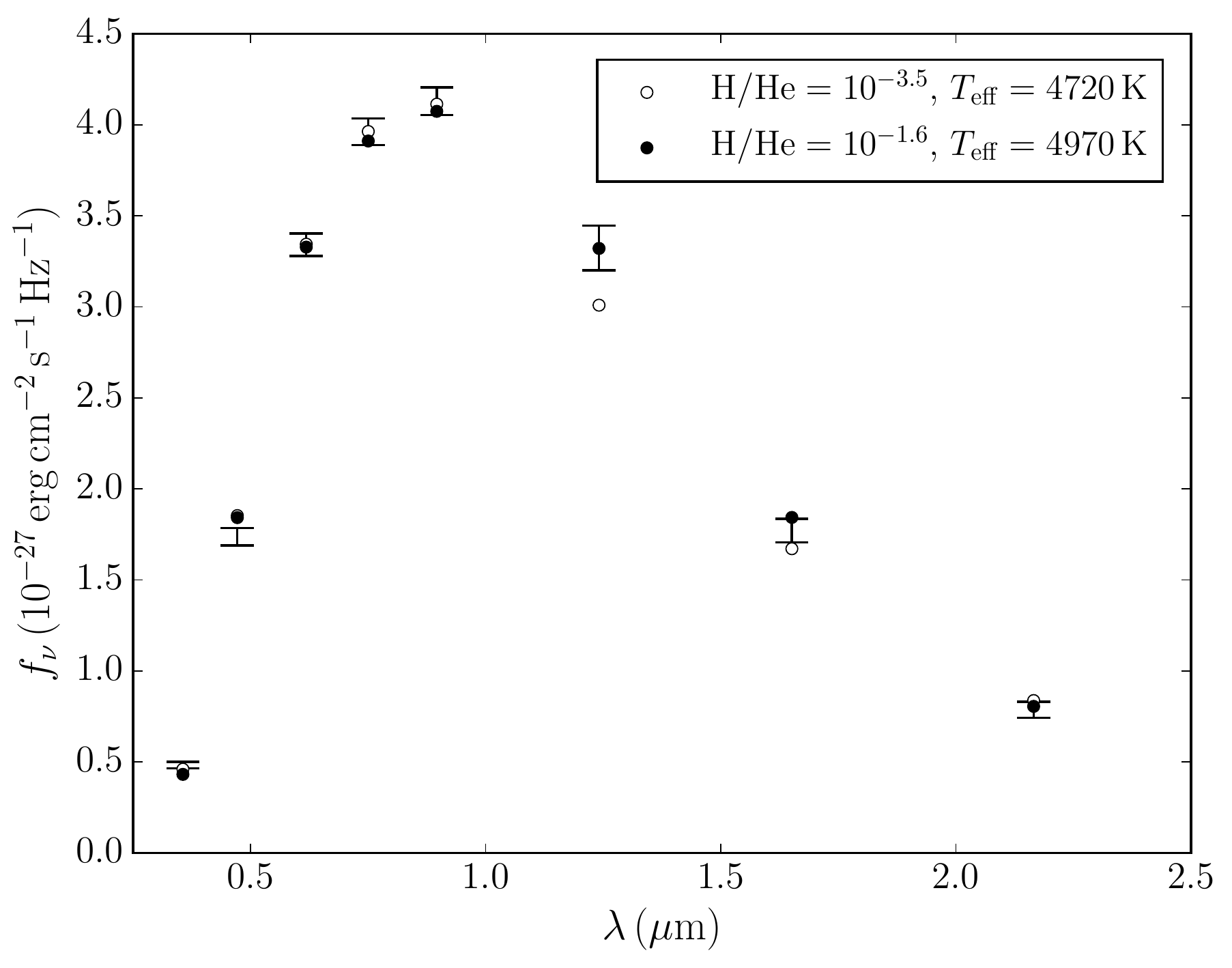}
  \caption{Best photometric solutions for models with $\log {\rm H/He}=-1.6$
    and $\log {\rm H/He}=-3.5$. The first solution
    has a metal abundance of $\log {\rm Ca/He}=-10.0$ and fits the spectroscopic data 
    very well (Figure \ref{fig:bestfit_spectro}), while
    the second solution has a metal abundance of $\log {\rm Ca/He}=-11.0$
    and is unable to reproduce the spectral lines observed
    in J0804+2239.}
  \label{fig:degen}
\end{figure}  

Can our analysis of J0804+2239 also be tainted by uncertainties associated with this degeneracy of the CIA intensity?
As in the examples given in the previous paragraph, we found that it is possible to obtain a second photometric solution for
J0804+2239 using a hydrogen abundance below the CIA maximum at $\log {\rm H/He} \approx -2.5$. This solution, found at
$\log {\rm H/He}=-3.5$, is shown in Figure \ref{fig:degen}. This second photometric solution is slightly better that our original
solution at $\log {\rm H/He}=-1.6$: the reduced $\chi^2$ is 1.9. 
However, in order to reach the right density conditions to obtain this second solution, 
we had to decrease the metal abundance down to $\log {\rm Ca/He}=-11.0$. 
Such a low metallicity is totally incompatible with the metal lines visible in the spectrum of J0804+2239. 
At $\log {\rm Ca/He}=-11.0$,
the spectral lines are simply too shallow to reproduce the observed spectrum, and hence this  
second solution can safely be discarded. We can therefore be confident that our fit of J0804+2239 is not affected 
by any degeneracy related to the CIA intensity since the spectroscopic data enables us to
lift this degeneracy. 
Note that if J0804+2239 had not shown any metal lines (as it is the case with DC stars), 
we could not have eliminated this degeneracy.

\subsection{DC stars with undetected metals}
In this section, we discuss of another type of degeneracy: undetectable metals in DC stars.
It is well known that the inclusion of metals in atmosphere models can significantly affect the effective temperature
derived from a photometric fit (compare for example the results found by \citealt{bergeron1997chemical,bergeron2001photometric} 
to those of \citealt{dufour2005detailed,dufour2007spectral} for the same DQ and DZ samples).
In this context, Figure \ref{fig:degen} raises an interesting question. 
Can a spectroscopically undetectable amount of metals create a degeneracy in the photometric solution 
of DC stars that show CIA features? If it is the case, the atmospheric parameters of many DC stars could be wrong
due to the impossibility of lifting this degeneracy.

\begin{figure}
  \includegraphics[width=\linewidth]{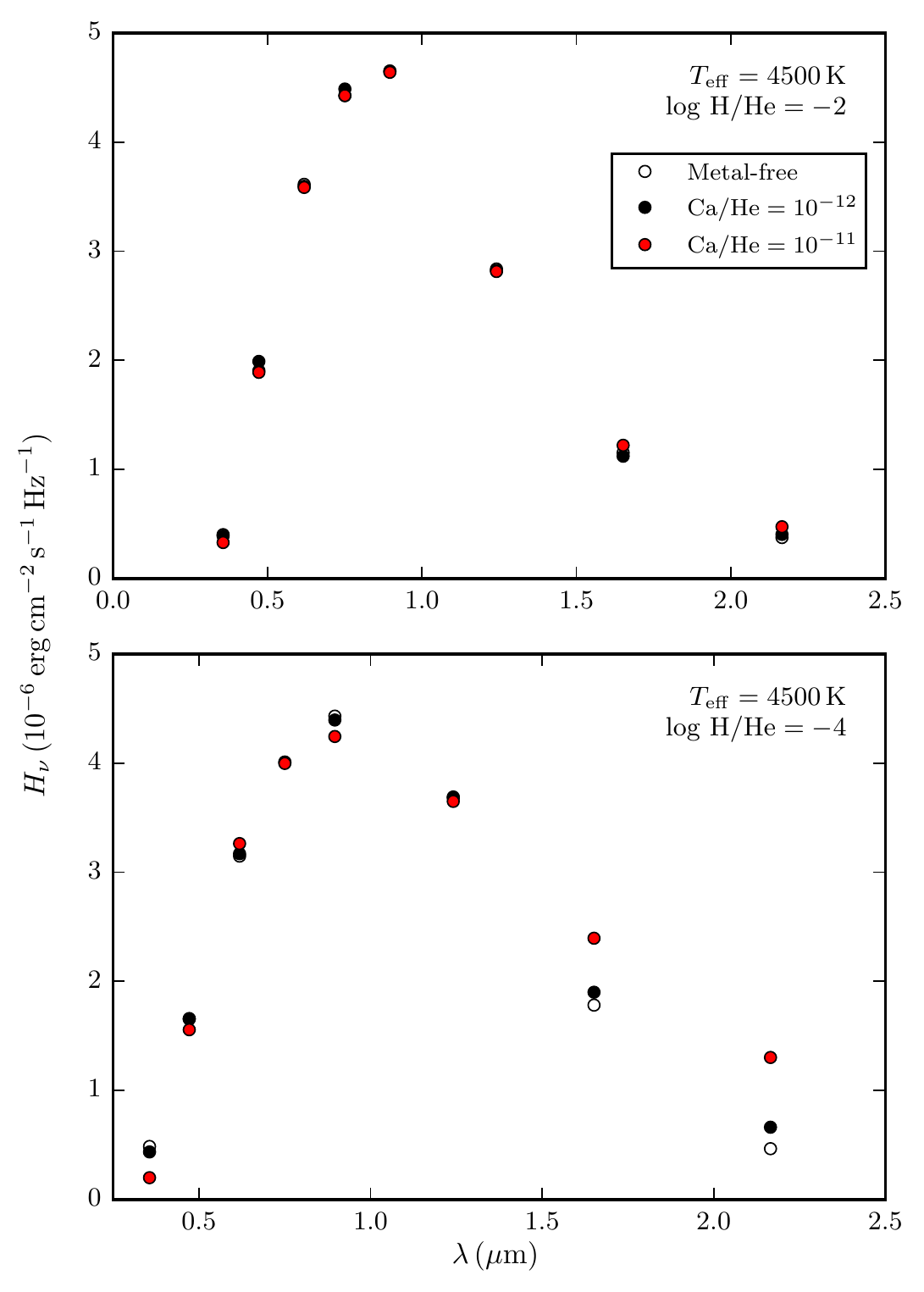}
  \caption{$ugriz$ and $JHK$ photometry of atmosphere models with $T_{\rm eff}=4500\,{\rm K}$ and $\log g =8$. The top panel
  shows SEDs of models with $\log {\rm H/He}=-2$ and the bottom one those of $\log {\rm H/He}=-4$ models.
  On each panel, we compare the SEDs of models with different metal abundances, as indicated in the legend.
  For the top panel, note that the metal-free SED is virtually identical to the SED with $\log {\rm Ca/He}=-12$.}
  \label{fig:DCphoto}
\end{figure}

To investigate this point, we compared $ugriz$ and $JHK$ photometry of atmosphere models that have the same
effective temperature and hydrogen abundance, but that have different metal abundances
(note that a surface gravity $\log g = 8$ is assumed throughout this section).
Figure \ref{fig:DCphoto} shows a sample of these comparisons for models without any heavy elements, with 
$\log {\rm Ca/He}=-12$ and with $\log {\rm Ca/He}=-11$. (For a typical spectrum with a signal-to-noise ratio of 
50 and a spectral resolution of 1\,{\AA}, the spectroscopic detection threshold of metals is
$\log {\rm Ca/He} \approx -11.5$, assuming $T_{\rm eff}$ and $\rm{H/He}$ values similar to the models of
Figure \ref{fig:DCphoto}.) We notice that the SEDs of models with a relatively high hydrogen abundance 
($\log {\rm H/He}=-2$) are barely affected by the addition of heavy elements, while the SEDs of models with a lower
hydrogen abundance ($\log {\rm H/He}=-4$) are much more affected. There are two reasons for this difference. First,
hydrogen is much less transparent than helium. As a result, for a model with little hydrogen, it does not take
a lot of heavy elements before we start "seeing" metals in the SED. Secondly, the atmosphere model structure
is less affected by the addition of metals for the $\log {\rm H/He}=-2$ models since hydrogen provides most
of the free electrons, which largely control continuum opacities and the photospheric pressure. 
For the $T_{\rm eff}=4500\,{\rm K}$,
$\log {\rm H/He}=-2$ and $\log {\rm Ca/He}=-11$ model, hydrogen atoms provide 96.9\% of the free electrons at the
photosphere, while heavy elements account for a mere 3.1\%. Since the addition of metals barely affects the free
electron density, it is unsurprising that it hardly affects the SED. On the opposite, for the
$T_{\rm eff}=4500\,{\rm K}$, $\log {\rm H/He}=-4$ and $\log {\rm Ca/He}=-11$ model, heavy elements provide
23.4\% of the free electrons at the photosphere and can therefore significatively influence the atmosphere model
structure.

From our analysis of Figure \ref{fig:DCphoto}, we can conclude that for stars with a relatively high hydrogen abundance
there is no danger of solution degeneracies caused by a spectroscopically undetectable amount of metals. However,
for DC stars with a smaller amount of hydrogen (e.g., $\log {\rm H/He}=-4$), the SED is changed even when a
spectroscopically undetectable amount of metals is added to the atmosphere and there is therefore a risk of
finding an erroneous photometric solution. In other words, it is a priori possible that
fitting a white dwarf that has a small quantity of metals (e.g., $\log {\rm Ca/He}=-12$) using metal-free models yields
an effective temperature and hydrogen abundance that are incorrect given that the influence of metals on the atmosphere 
structure of this hypothetical star was not taken into account.

If this scenario has a chance to come true, then it should be possible for a metal-free
model to emulate the SED of a model with a small quantity of metals and a different $T_{\rm eff}$ and ${\rm H/He}$ 
abundance ratio.
To test if it is the case, we tried to fit the SEDs of cool DZ stars using a grid of helium-rich DC models.
For these fits, both the effective temperature and the H/He ratio were adjusted. We explored a parameter space
extending from $T_{\rm eff}=4000$ to $5500\,{\rm K}$, from $\log {\rm H/He}=-6$ to $-3$, and from
$\log {\rm Ca/He} = -13$ (well below the detection threshold) to $-11$ (slightly above the detection threshold).
In the worst case, we find that the effective temperature obtained by the fitting procedure has a 
250\,K discrepancy with the actual temperature of the DZ model that is being fitted. 
More important are the discrepancies found on H/He (which can reach 3 dex), but this is simply a manifestation
of the CIA degeneracy described in Section \ref{sec:cia_degen} and not an effect attributable to the
presence of metals in the atmosphere.
Therefore, we can conclude that the degeneracy due to the presence of undetected metals does not generate 
errors that go beyond the usual fitting uncertainties.

\begin{figure}
  \includegraphics[width=\linewidth]{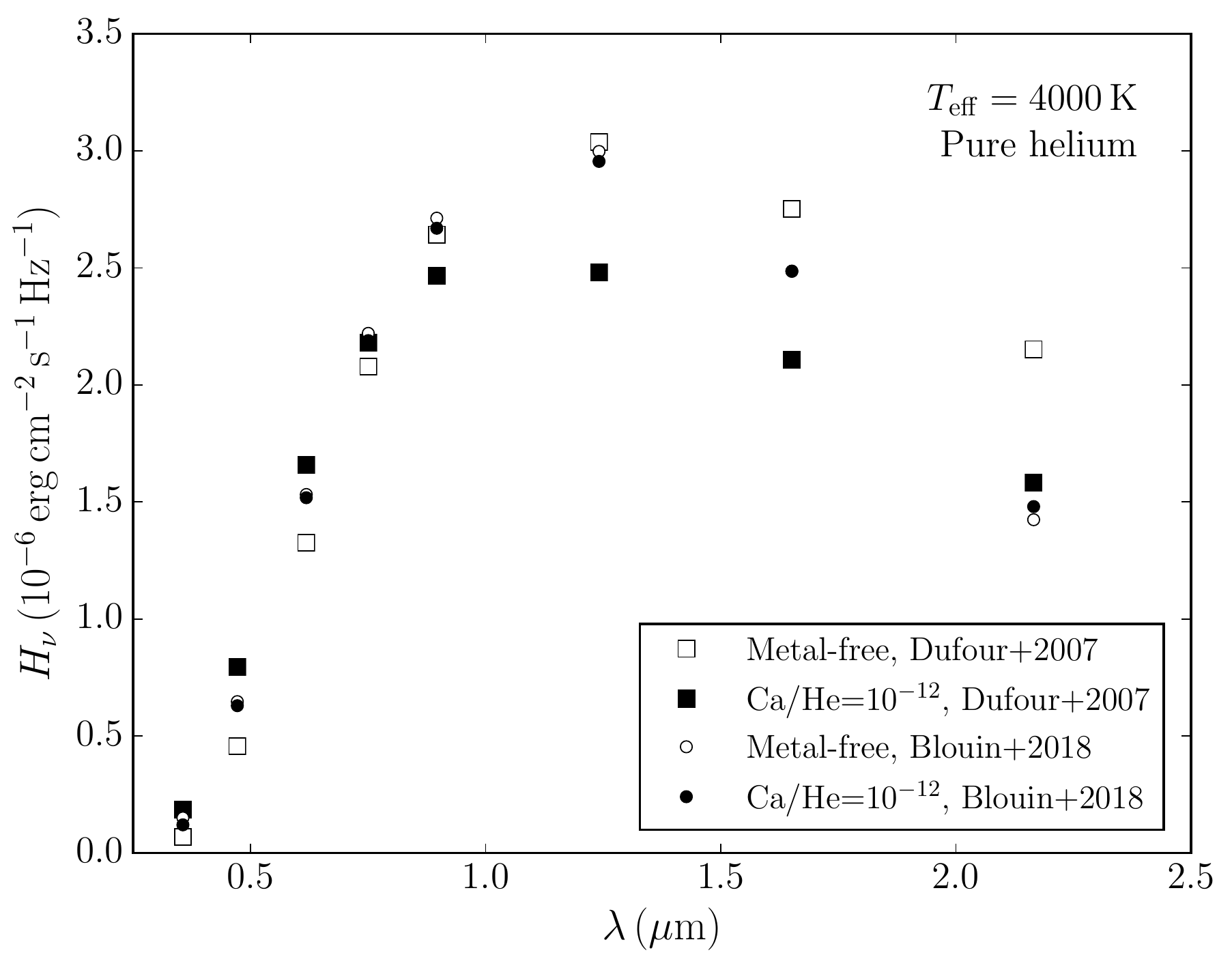}
  \caption{SEDs of hydrogen-free atmosphere models with $T_{\rm eff}=4000\,{\rm K}$ and $\log g =8$. The circles
    correspond to the results obtained using the models of \cite{blouin2018model} and the squares to those
    found with the models of \cite{dufour2007spectral}. For the later, note how the SED changes when a
    minute amount of metals is added to the atmosphere.
  }
  \label{fig:old_vs_new}
\end{figure}

Note that these conclusions also extend to DC stars that do not show any CIA feature ($\log {\rm H/He} < -5$).
This result is quite different than what was previously found using the models of \cite{dufour2007spectral}.
Figure \ref{fig:old_vs_new} shows how the SED of those models are affected by the addition of a small amount of metals.
At $T_{\rm eff}=4000\,{\rm K}$, a metal abundance as low as $\log {\rm Ca/He}=-12$ leads to drastic differences
in the shape of the SED. As shown in Figure \ref{fig:old_vs_new}, such a difference is not visible
with our improved models. The main explanation for the disagreement between both sets of models
lies in the atmosphere model structure of pure-helium models. In particular, as the pressure ionization
of helium \citep{kowalski2007equation} was not implemented in the models of \cite{dufour2007spectral},
their pure-helium models are much more transparent than our new models.
The photospheric pressure reaches $10^{13}\,{\rm dyn/cm}^2$ for $T_{\rm eff}=4000\,{\rm K}$, 
more than 20 times the pressure of a model with $\log {\rm Ca/He}=-12$. On the contrary, with our
improved models, the model atmosphere structure without metals and with $\log {\rm Ca/He}=-12$ are
very similar. Because of the pressure ionization of helium, free electrons are still abundant when metals
are absent. For that reason, the intensity of He$^-$ free-free absorption does not fall dramatically as it is
the case with the models of \cite{dufour2007spectral}.

\section{Conclusion}
\label{sec:conclusion}

We presented the first spectroscopic and photometric fit of J0804+2239, a unique cool
DZ star that shows CIA. The improvements recently made to our atmosphere code
(in particular, the improved high-pressure Ca line profiles and the new H$_2$-He CIA profiles)
proved to be crucial to obtain a good fit. These results show that our upgraded
code properly captures the physics of the moderately dense atmospheres of cool DZ stars.

Because it contains metals, the hydrogen abundance in the atmosphere of J0804+2239
could be determined with greater certainty than the hydrogen abundance of most white dwarfs
with CIA features. In fact, the presence of metal absorption lines lifts the degeneracy
between high and low hydrogen abundances.
We also explored the possibility that spectroscopically undected metals could affect
the photometric solutions of DC stars. We found that in the worst case the errors
induced on the atmospheric parameters are of the order of the usual fit uncertainties.

In the next paper of this series, we will continue the observational validation of
our model atmosphere code with still cooler DZ stars. In particular, we will revisit
WD 2356-209, a cool DZ with an atypically strong Na D doublet.

\acknowledgments

This work was supported in part by NSERC (Canada).
This work has made use of the Montreal White Dwarf Database \citep{dufour2016montreal}.

This work has made use of data from the European Space Agency (ESA) mission
{\it Gaia} (\url{https://www.cosmos.esa.int/gaia}), processed by the {\it Gaia}
Data Processing and Analysis Consortium (DPAC,
\url{https://www.cosmos.esa.int/web/gaia/dpac/consortium}). Funding for the DPAC
has been provided by national institutions, in particular the institutions
participating in the {\it Gaia} Multilateral Agreement.\\

\bibliographystyle{aasjournal}
\bibliography{references}

\end{document}